\documentclass[prl,twocolumn,superscriptaddress]{revtex4-1}

\usepackage{amsmath}
\usepackage[utf8]{inputenc}
\usepackage{braket}
\usepackage{float}
\usepackage{mathtools}

\usepackage{pgfplots}

\usepackage{hyperref}
\usepackage{lipsum}
\usepackage{xcolor}
\usepackage{comment}
\usepackage{bm,amssymb,graphicx,braket,dsfont,mathtools,bbold,MnSymbol}

\DeclareMathOperator{\sign}{sign}

\usepackage[caption=false]{subfig}
\usepackage{pgfplots}
\usepackage{cancel}
\usepackage{tabularx,booktabs}
\usepackage{multirow}
\usepackage{fancyhdr}
\usepackage{amsthm}
\usepackage{inputenc}
\usepackage{soul}
\usepackage{nccmath}

\newcommand{\vs}{\vec{\sigma}}
\newcommand{\tr}{\mathrm{Tr}}
\newcommand{\lind}{\mathcal{L}}
\newcommand{\proj}[1]{\ket{#1} \bra{#1}}
\newcommand{\abs}[1]{\left| #1 \right|}
\newcommand{\V}{\mathcal{V}}

\newcommand{\dgm}{\mathcal{D}_{\gamma_{\mu}}}
\newcommand{\dkm}{\mathcal{D}_{\kappa_{\mu}}}

\begin{document}

\title{Signatures of associative memory behavior in a multi-mode spin-boson model}
\author{Eliana Fiorelli}
\address{School of Physics and Astronomy, University of Nottingham, Nottingham, NG7 2RD, UK}
\address{Centre for the Mathematics and Theoretical Physics of Quantum Non-equilibrium Systems, University of Nottingham, Nottingham NG7 2RD, UK} 
\author{Matteo Marcuzzi}
\address{School of Physics and Astronomy, University of Nottingham, Nottingham, NG7 2RD, UK}
\address{Centre for the Mathematics and Theoretical Physics of Quantum Non-equilibrium Systems, University of Nottingham, Nottingham NG7 2RD, UK} 
\author{Pietro Rotondo}
\address{School of Physics and Astronomy, University of Nottingham, Nottingham, NG7 2RD, UK}
\address{Centre for the Mathematics and Theoretical Physics of Quantum Non-equilibrium Systems, University of Nottingham, Nottingham NG7 2RD, UK} 
\author{Federico Carollo}
\address{School of Physics and Astronomy, University of Nottingham, Nottingham, NG7 2RD, UK}
\address{Centre for the Mathematics and Theoretical Physics of Quantum Non-equilibrium Systems, University of Nottingham, Nottingham NG7 2RD, UK} 
\address{Institut für Theoretische Physik, Universität Tübingen, Auf der Morgenstelle 14, 72076 Tübingen, Germany}
\author{Igor Lesanovsky}
\address{School of Physics and Astronomy, University of Nottingham, Nottingham, NG7 2RD, UK}
\address{Centre for the Mathematics and Theoretical Physics of Quantum Non-equilibrium Systems, University of Nottingham, Nottingham NG7 2RD, UK}
\address{Institut für Theoretische Physik, Universität Tübingen, Auf der Morgenstelle 14, 72076 Tübingen, Germany}
\date{\today}

\begin{abstract} 
Spin-boson models can describe a variety of physical systems, such as atoms in a cavity or vibrating ion chains. In equilibrium these systems often feature a radical change in their behavior when switching from weak to strong spin-boson interaction. This usually manifests in a transition from a ``dark" to a ``superradiant" phase. However, understanding the out-of-equilibrium physics of these models is extremely challenging, and even more so for strong spin-boson coupling. Here we show that non-equilibrium strongly interacting spin-boson systems can mimic some fundamental properties of an associative memory - a system which permits the recognition of patterns, such as letters of an alphabet. Patterns are encoded in the couplings between spins and bosons, and we discuss the dynamics of the spins from the perspective of pattern retrieval in associative memory models. We identify two phases,  a ``paramagnetic" and a ``ferromagnetic" one, and a crossover behavior between these regimes. The ``ferromagnetic" phase is reminiscent of pattern retrieval. We highlight similarities and differences with the thermal dynamics of a Hopfield associative memory and show that indeed elements of ``machine learning behavior" emerge in strongly coupled spin-boson systems.

\end{abstract}

\maketitle
%%%
\emph{Introduction---}
Cavity, circuit \cite{Blais:PRA:2004} and waveguide quantum electrodynamics (QED)  \cite{Zheng:PRL:2013}, as well as trapped ions \cite{Cirac:PRL:1995,Leibfried:RMP:2003} provide controllable platforms for quantum simulation. These systems are often described as collections of two-level components (spins) coupled via one or more bosonic degrees of freedom (such as photons or phonons). At the most fundamental level, their properties are captured by the Dicke model \cite{Garraway11,Kirton:AQT:2019} which, under equilibrium conditions and for large number of spins $N \to \infty$, displays a quantum phase transition from a dark to a superradiant phase: for weak spin-boson coupling the average occupation number of bosons is sub-extensive (it grows slower than $N$ with the number of spins), whereas it is extensive ($\propto N$) above a critical threshold \cite{HeppL73,HeppL:PRA:73}. This phase transition has been widely investigated \cite{Emary:PRE:2003,Emary:PRL:2003,Nagy:PRL:10}, observed in Bose-Einstein condensates in optical cavities \cite{Baumann:PRL:10, Baumann:Nat:10} and recently studied in more general, non-equilibrium settings \cite{Kirton:PRL:2017,Gambetta19}.

While originally predicted for a single bosonic degree of freedom, this phase transition was eventually generalized to a multi-mode scenario \cite{Gopalakrishnan:NatPhys:09}, where each spin interacts with several different bosonic modes. Interesting phenomena can emerge in this setting, in particular when the spin-boson couplings are non-uniform, and depend on both the specific spin and the specific boson \cite{MoralesEtAl19, Guo:PRL:2019}. In this case the superradiant phase features glassy behavior \cite{Goldbart:PRL:2011, Strack:PRL:2011,Rotondo:PRB:2015} as well as an energy landscape that shares similarities with simple neural networks \cite{Rotondo:PRL:2015, Gopalakrishnan:phil, Torggler2017, Damanet:PRA:2019}.

\begin{figure}[h!]
\includegraphics[width=\linewidth]{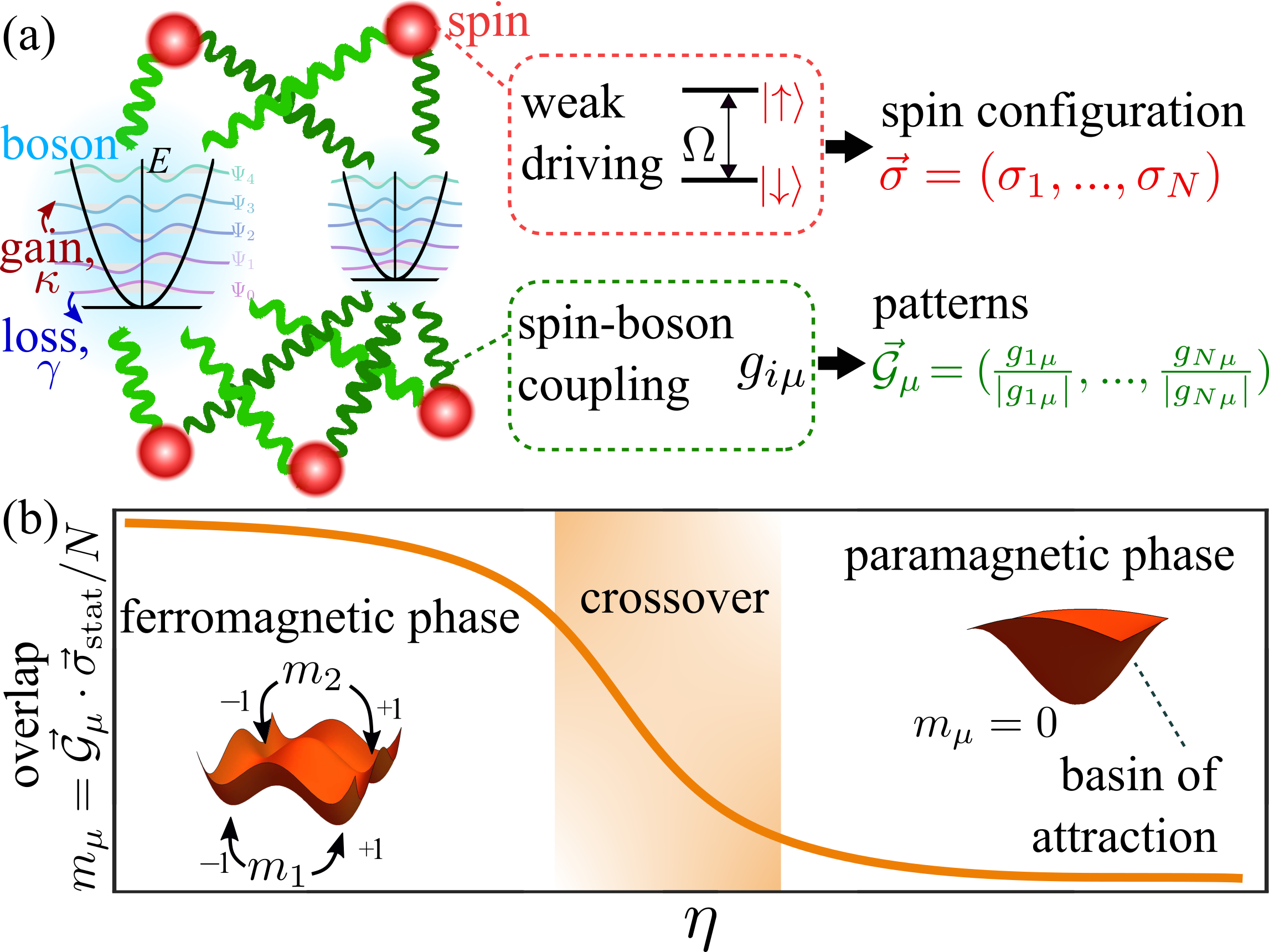}
\caption{\textbf{Disordered, dissipative spin-boson system.} (a) $N$ spin-$1/2$ particles are strongly coupled to two bosonic modes with couplings $g_{i \mu}$ ($i=1,...,N$, $\mu=1,2$). The spins are weakly driven at a strength $\Omega \ll g_{i \mu}$. Each bosonic mode is subjected to two dissipative processes, gain and loss, occuring at rates $\kappa$ and $\gamma$ respectively. Patterns are encoded in the couplings between the spins and the bosons, corresponding to $\vec{\mathcal{G}}_{\mu}=[\sign(g_{1 \mu}),..., \sign(g_{N \mu})]^{T}$ ($\mu=1,2$).  (b) Phase diagram of the overlap $m_{\mu}=(\mathcal{G}_{\mu} \cdot \vs_{\mathrm{stat}} )/N$ between the classical, stationary spin configuration and the pattern as a function of the  parameter $\eta=(\gamma-\kappa)/\omega$ (for details see text). A smooth transition between two phases occurs, these being characterized by the presence of several (ferromagnetic or ``retrieval" phase) and one (paramagnetic one) basins of attraction. The insets show sketches of the effective free energy landscape.}\label{fig.intro}
\end{figure}

%%%%%%%%%%%%%%%%%%%%%%%%%%%%%%%%%%%%%%%%%%%%%%%
%In this work we investigate the connection between such multimode spin-boson systems and the dynamics of an associative memory, the Hopfield neural network (HNN) \cite{Hopfield:1982, Amit_book}, in an out-of-equilibrium setting. In particular, instead of considering a closed system at equilibrium we couple it with an external environment. It is known that the superradiant phase can survive the introduction of dissipation \cite{Kirton:PRL:2017}. The purpose of this work is to assess how robust the associative memory features are in an open setting. Our analysis shows that retrieval of information is still possible to a degree, depending on the choice of the parameters, suggesting that memory effects can also be investigated in driven-dissipative settings. 
%
%%%%%%%%%%%%%%%%%%%%%%%%%%%%%%%%%%%%%%%%%%%%%%%%%%
%
In this work, we consider a driven-dissipative multi-mode spin-boson model, sketched in Fig.~\ref{fig.intro}(a), and show that its stationary properties resemble those of the so-called Hopfield neural network (HNN) \cite{Hopfield:1982, Amit_book}, a classical spin model behaving like a basic associative memory. More precisely, we identify a crossover between a disordered ``paramagnetic" phase and an ordered ``ferromagnetic" one [see Fig.~\ref{fig.intro}(b)]; the latter appears to be closely related to the retrieval phase of the HNN, characterised in turn by the ability to recall previously stored information. We furthermore explore the impact of fluctuating spin-boson coupling constants on this crossover. 
Our results show that analogies between multi-mode spin-boson systems and phases of the HNN are not limited to equilibrium settings \cite{Rotondo:PRB:2015}.

\emph{The Hopfield model: a brief recap---} In physical terms, the HNN is a fully-connected classical spin model consisting of $N$ Ising variables (``neurons") $\sigma_i = \pm 1$, $i = 1 \ldots N$, collated to form configurations $\vec{\sigma} = \left( \sigma_1, \ldots , \sigma_N \right)^T$. Information is stored in the form of a number $M$ of special configurations, or ``patterns", $\vec{\xi_{\mu}}$, $\mu = 1 \ldots M$. In what follows, we employ Roman (Greek) letters for site (pattern) indices. The patterns enter the definition of the energy function
\begin{equation}\label{eq.Hopfield} 
 E_{\mathrm{Hop}}(\vec{\sigma})= - \frac{1}{2 } \sum_{i\neq j}^{N}J_{i j}\sigma_{i}\sigma_{j },
\end{equation}
via the connectivity matrix $J_{ij}=\frac{1}{N}\sum_{\mu=1}^{M} \xi_{i \mu} \xi_{j \mu}$. Under the hypotheses that (i) the patterns are ``few'' $M / N < 0.14$ and (ii) different patterns are approximately orthogonal, or more precisely, that they are chosen in such a way that $\lim_{N \to \infty} \left( \vec{\xi}_\mu \cdot \vec{\xi}_\nu / N \right) = \delta_{\mu \nu}$, one can prove that the energy is minimised by the $2M$ configurations $\vec{\sigma} = \pm \vec{\xi_{\mu}}$ ($\mu = 1 \ldots M$) \cite{Amit:PRL:1985}. In other words, the patterns (and their opposites $-\vec{\xi_{\mu}}$) are the ground states of the system and an ideal annealing procedure to zero temperature would recover them perfectly. In order to quantify such retrieval, it is convenient to introduce the $M$ overlaps $\zeta_\mu = \vec{\xi}_\mu \cdot \vec{\sigma} /N$ which measure how much the configuration aligns with any given stored pattern. For sufficiently large $N$, a frequent choice is to describe the pattern components $\xi_{i\mu}$ as independent random variables which take the values $\pm 1$ with equal probabilities. 

Similarly to a fully-connected Ising model, the HNN undergoes a continuous equilibrium phase transition at inverse temperature $\beta = 1$, from a paramagnetic phase ($\beta < 1$) in which $\zeta_\mu \to 0 \,\, \forall \mu$ to a retrieval (or ferromagnetic) phase $\beta > 1$ where a single component $\zeta_{\bar{\mu}}$ acquires a non-vanishing value, while $\zeta_\mu \to 0 \,\, \forall \mu \neq \bar{\mu}$. 
In the retrieval phase (and in the thermodynamic limit), a thermal dynamics (e.g. heat-bath) implemented on the HNN will eventually lead to the configuration approximately reproducing the features of one of the stored patterns. The specific choice can be dictated by the initial condition: an initial configuration with small, but non-negligible, overlap $\zeta_{\bar{\mu}}$ will restrict the dynamics to the corresponding basin of attraction and the configuration at long times will fluctuate approximately around $\pm \vec{\xi}_{\bar{\mu}}$. This ability to faithfully, up to some noise, reconstruct a pattern from partial initial information is what qualifies the HNN as an associative memory. 

%We refer the reader elsewhere \cite{Hopfield:1982, Amit_book} for a more in-depth description of the HNN phases, which is richer than presented here, but beyond the scope of this work.
%
%We briefly remark that the Hopfield physics is actually richer than presented above and includes the emergence of metastable states at very low temperatures and a spin glass phase when one tries to store patterns beyond the system's capacity ($M / N > 0.14$). A comprehensive review is beyond the present scope; we refer the interested reader elsewhere \cite{Hopfield:1982, Amit_book}.

\emph{Non-equilibrium spin-boson model---} 
We now introduce the open quantum spin-boson system of interest. As sketched in Fig.~\ref{fig.intro}(a), we consider $N$ spin-$1/2$ particles interacting with $M$ independent bosonic modes, according to the Dicke-like \cite{Garraway11,Kirton:AQT:2019,HeppL73,HeppL:PRA:73} Hamiltonian 
$\hat{H}=\sum_{\mu=1}^{M}\omega_{\mu}\hat{a}_{\mu}^{\dagger}\hat{a}_{\mu}+\sum_{\mu=1}^{M}\sum_{i=1}^{N}g_{i \mu}\hat{\sigma}_{i}^{z}(\hat{a}_{\mu}^{\dagger}+\hat{a}_{\mu})+ \Omega\sum_{i=1}^{N}\hat{\sigma}_{i}^{x}\,
$. Here, $\hat{\sigma}_{i}^{x,y,z}$ are the $i$-th spin Pauli operators, $\hat{a}_{\mu}$ and $\hat{a}_{\mu}^{\dagger}$ the annihilation and creation operators of the $\mu$-th bosonic mode, $\omega_{\mu}$ the corresponding frequency. The parameter $\Omega$ drives transitions between spin states and the $g_{i\mu}$s are the spin-boson couplings. For concreteness, we assume these to be independent, identically-distributed, real random variables.
%  with vanishing mean and variance $g^2$. 

Additionally, the system exchanges bosons with a Markovian bath. The system state $\rho$ evolves according to a Lindblad equation \cite{BreuerP:2002, Lindblad76} 
$
\dot{\rho}=\lind \rho=-i [\hat{H}, \rho] +\sum_{\mu,n=l,g} \hat{L}_{n,\mu}\rho \hat{L}_{n,\mu}^{\dagger}-\frac{1}{2} \lbrace \hat{L}_{n,\mu}^{\dagger} \hat{L}_{n,\mu}, \rho \rbrace,
$
where jump operators $\hat{L}_{l, \mu}=\sqrt{\gamma_{\mu}}\hat{a}_{\mu}, ~ \hat{L}_{g, \mu}=\sqrt{\kappa_{\mu}}\hat{a}_{\mu}^{\dagger}$ describe independent processes of loss and gain of bosons at rates $\gamma_{\mu} > \kappa_{\mu} \geqslant 0$. 
%
%The sketch in Fig. \ref{fig.intro}~(a) shows a system with $\gamma_{\mu}=\gamma$ and $\kappa_{\mu}=\kappa$ $\forall \mu = 1,...,M$, a restriction we also employ for simplicity in our simulations further below.

As for the case of a single boson \cite{PhysRevResearch.2.013198}, we can obtain an effective master equation restricted to the spin degrees of freedom only. 
%
%
%To this end we call ``classical configurations'' $\ket{\vec{\sigma}} = \ket{(\sigma_1 , \ldots , \sigma_N)^T}$ the eigenstates of the $\hat{\sigma}^z_i$  spin operators, i.e., $\hat{\sigma}^z_i \ket{\vec{\sigma}} = \sigma_i \ket{\vec{\sigma}}$. Our aim is to obtain an effective dynamics for the spins only. To this end, we generalize a method introduced in \changer{[ref]} for a single boson to the present multimode case. 
% 
The technical details are reported in the Supplementary Material \cite{SM}; here we summarize the main conceptual steps: for vanishing $\Omega$, each $\hat{\sigma}^z_i$ is a conserved quantity; it is therefore natural to focus our attention on the $z$-component eigenbasis $\ket{\vec{\sigma}} = \ket{(\sigma_1 , \ldots , \sigma_N)^T}$, $\hat{\sigma}^z_i \ket{\vec{\sigma}} = \sigma_i \ket{\vec{\sigma}}$, which is defined in terms of ``classical configurations'' $\vec{\sigma}$. Due to their conservation, each configuration labels, at $\Omega = 0$, a subspace of states disconnected from the others, where the only non-trivial evolution takes place in the bosonic part. In particular, for any fixed $\ket{\vec{\sigma}}$ the bosonic stationary state is a displaced Gaussian state $\rho_{\vec{\sigma}}$. This implies that the dynamics has a degenerate stationary space spanned by the $2^N$ elementary combinations $\proj{\vec{\sigma}} \otimes \rho_{\vec{\sigma}}$. This ``classical'' subspace gets dynamically coupled to the remainder of the space by the introduction of the term $\propto \Omega$; however, as long as $\Omega$ is sufficiently small, standard perturbative techniques \cite{Nakajima58,zwanzig1960ensemble} can be employed to project the resulting dynamics back onto the classical subspace. By additionally tracing over the bosonic modes, a reduced spin dynamics is found \cite{Degenfeld14, Marcuzzi:JPA:2014}. Up to order $\Omega^2$, this is encoded in a classical rate equation
\begin{equation}\label{eq.Stochastic}
\dot{p}_{\vs}= \sum_{\vs'} \left( W_{\vs' \rightarrow \vs}p_{\vs'}-W_{\vs \rightarrow \vs'}p_{\vs} \right),
\end{equation}
with $p_{\vec{\sigma}}$ the probability to find the spins in the configuration $\vec{\sigma}$ and $W_{\vs \rightarrow \vs'}$ the transition rate for switching from configuration $\vs$ to $\vs'$. Due to the structure of the perturbative term, the only allowed elementary processes are single spin flips, i.e. $W_{\vs \rightarrow \vs'} \neq 0$ exclusively when $\vs$ and $\vs'$ differ by a single spin. Restricting for simplicity to $\gamma_{\mu}\equiv \gamma$, $\kappa_{\mu}\equiv \kappa$, $\omega_{\mu}\equiv \omega$, $\forall \mu$, the derivation of Eq.~\eqref{eq.Stochastic} can be found in Ref. \cite{SM}. We also set $\omega = 1$.
%, i.e. all quantities are measured in units of $\omega$.
%Furthermore, the detailed balance condition \cite{Zia_2007} on the transition rates does not hold, hence the effective spin dynamics is in general out-of-equilibrium. 

\emph{Energy function, overlaps and coupling distribution---}
Remarkably, transition rates only depend on the spin configuration through the quantity 
$\Delta E_{i}= \sigma_i \sum_{\mu, j \neq i} g_{i \mu} g_{j \mu} \sigma_{j}$, which can be regarded as the cost $\Delta E_{i}=E(-\sigma_{i})-E(\sigma_{i})$ of flipping the $i$-th spin given an energy function of the form
\begin{equation}\label{eq.all-Ising}
E(\vs)=-\frac{1}{4}\sum_{\mu}\sum_{i \neq j} g_{i \mu} g_{j \mu} \sigma_{i } \sigma_{j }.
\end{equation}
Equation \eqref{eq.all-Ising} bears a clear resemblance to the Hopfield energy \eqref{eq.Hopfield} if one replaces the patterns $\xi_{i \mu}$ with the spin-boson couplings $g_{i \mu}$. With this analogy in mind, we  interpret the spin-boson couplings as ``noisy patterns''. Specifically, we pick the coupling costants $g_{i\mu}$ from a bimodal distribution peaked around $\pm 1$, $\mathcal{P}(g)=\frac{1}{2}[\mathcal{N}_{+1,s}(g)  +\mathcal{N}_{-1,s}(g)]$, given by the symmetric superposition of two Gaussians $\mathcal{N}_{g_{0},s}(g)=(2\pi s^2)^{1/2}\exp{-(g-g_{0})^{2}/(2s^{2})}$.  From this perspective, it is natural to introduce, as prospective order parameters, and in analogy to the HNN, the \emph{overlaps} $m_{\mu}=\sum_{i=1}^{N} \mathrm{sign}( g_{i \mu}) \sigma_{i}/N \equiv \vec{\mathcal{G}}_{\mu} \cdot \vs /N$, where $\vec{\mathcal{G}}_{\mu}= [\mathrm{sign}(g_{1 \mu}), ..., \mathrm{sign}(g_{N \mu})]^{T}$ represent the noiseless patterns.

For the sake of simplicity, we focus on  the behaviour of $m_\mu$ upon varying only the parameter $
\eta= \gamma - \kappa $ that quantifies the net bosonic loss rate. As we show further below, this parameter controls, to some extent, effective thermal fluctuations. We further fix $\kappa/\gamma=0.9$, $N=50$, $M=2$. Finally, as transition rates are $\propto \Omega^2$, we rescale the time accordingly and effectively set $\Omega = 1$.

\begin{figure*}
\includegraphics[width=\linewidth]{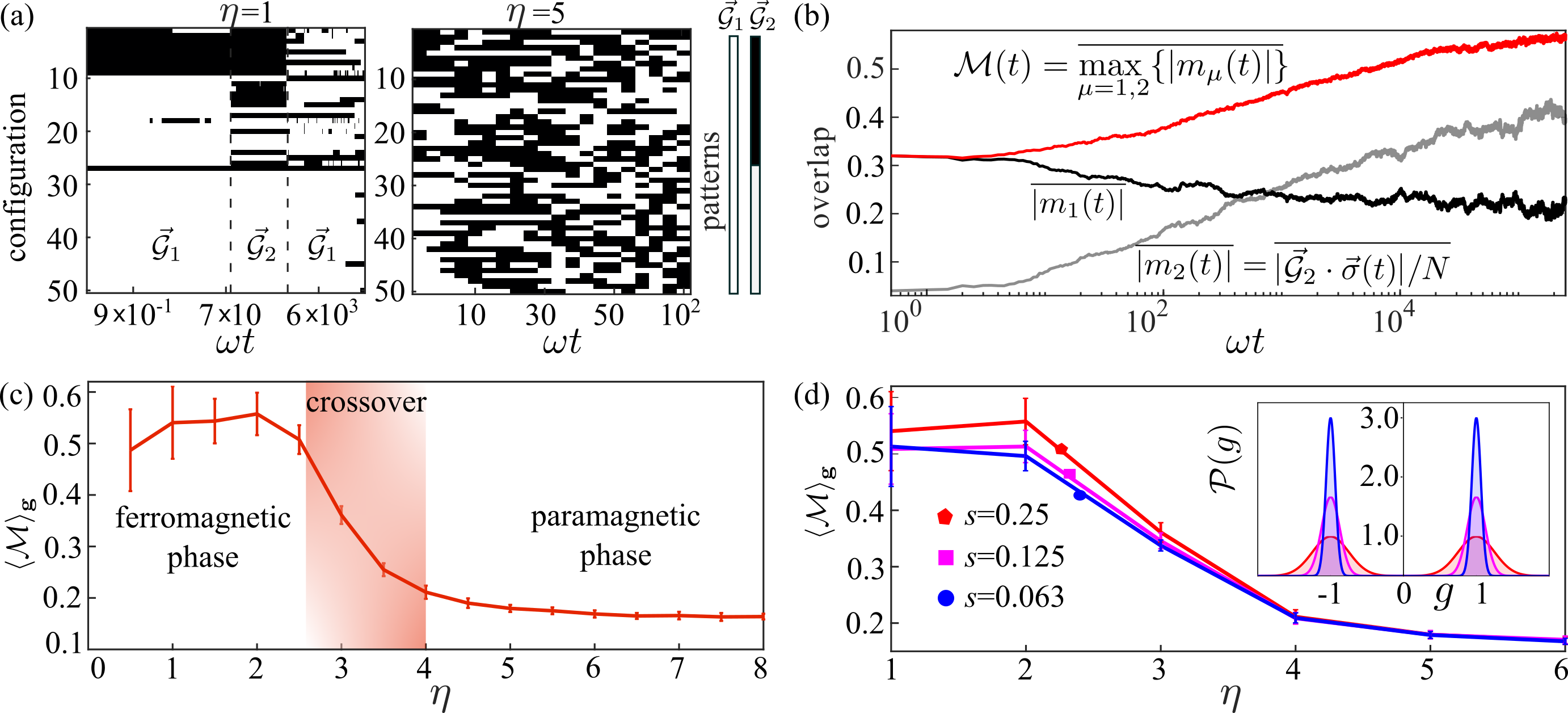}
\caption{(a)-(b) \textbf{Stochastic dynamics.} We select the spin-boson couplings from a bimodal distribution given by two superposed Gaussians centerd at $g_{0}= \pm 1$ and standard deviation $s=0.25$. (a) Time evolution of a spin configuration at fixed disorder realization. The two patterns $\vec{\mathcal{G}}_{1,2}$ are represented in the legend. At the initial time, the overlap between the spin configuration and $\mathcal{G}_{1}$ is $m_{1}(0)=0.6$. The left-hand side panel shows the time evolution at $\eta =1$, whereas the right-hand side shows the time evolution at $\eta =5$. (b) Time evolution of order parameter and overlaps as a function of time, at $\eta=1$. The overlaps are shown in terms of the absolute values averaged over $N_{\mathrm{traj}}=200$ realizations of the stochastic process \eqref{eq.Stochastic}. The initial configuration is chosen randomly. (c)-(d) \textbf{Stationary properties.} Disorder-averaged order parameter  $\braket{ \mathcal{M}}_{\mathbf{g}}$ as a function of $\eta$ and its standard deviation (half error bars)  $\pm ( \braket{\mathcal{M}^2}_{\mathbf{g}} - \braket{\mathcal{M}}_{\mathbf{g}}^2)^{1/2}$, resulting after the average over $N_{\mathrm{distr}}=30$ disorder realizations of the (symmetrized) order parameter $\mathcal{M}$. The specific times at which the points are taken differ at different $\eta$ since, as can be gleaned from Fig.~\ref{fig.stationary}(a), the dynamical time-scales differ greatly in the opposite regimes of small and large $\eta$. Lines are guides for the eyes. (c) The interaction coupling are selected from a bimodal distribution given by superposing two Gaussians centred in $g_{0}=\pm 1 $ and with standard deviation $s=0.25$. (d) Plots of the symmetrised order parameter against $\eta$ for three values of the standard deviation $s=0.25$ (red star), $s=0.125$ (magenta square), and $s=0.063$ (blue circle), as reported in the legend. Different Gaussians distribution are represented in the inset. Other parameters are $N=50$, $M=2$, $\omega=1$, $\kappa/\gamma=0.9$.
}\label{fig.stationary}
\end{figure*}

\emph{Non-equilibrium dynamics---} We simulate the dynamics (\ref{eq.Stochastic}) via kinetic Monte Carlo methods \cite{BortzKL1975, YoungE1966, Gillespie1976}. For convenience, we call different realizations of the stochastic process ``trajectories'', reserving ``realizations'' for different random choices of the couplings.

As a starting point of the analysis we consider that the HNN dynamics within the retrieval [paramagnetic] phase is characterized by the presence of multiple [a single] basins of attractions $\zeta_{\mu} \approx \pm 1$ [$\zeta_{\mu} \approx 0$]. 
%In order to understand if any of this features occurs in our model,  we concentrate the analysis of the effective spin model dynamics upon the spin configuration and  the overlaps $m_{\mu}$.
%
%
Two trajectories (at fixed coupling realization) of the stochastic process \eqref{eq.Stochastic} are shown in Fig.~\ref{fig.stationary}(a) for two distinct $\eta$ regimes. In both cases, the initial configuration is $80\%$ aligned with the first pattern, implying $m_1(t=0) = 0.6$. For ease of visualisation we apply a gauge transformation $\sigma_i \to \mathcal{G}_{i,1} \sigma_i$, $\mathcal{G}_{i,\mu} \to \mathcal{G}_{i,1} \mathcal{G}_{i,\mu} $ which aligns all the components of the first pattern. The components are then re-ordered to bring all the positive components of the second pattern on one side and all the negative ones on the other.  
In the small-$\eta$ regime (left-hand side) the dynamics is trapped for long times in configurations close to either pattern; these long-lasting periods are separated by fast switching from one pattern to the other, a behavior commonly seen in finite-size versions of systems undergoing spontaneous symmetry breaking. Phenomenologically [see Fig.~\ref{fig.intro}(b)], the free energy landscape over the space of configurations breaks into different, approximate basins of attraction (BA); the dynamics tends to remain confined in one such potential well until a rare, sufficiently large fluctuations is able to overcome the barrier, after which the dynamics is trapped again for long times in the new BA. A similar behavior would also be observed, at finite size, in the thermal dynamics of the HNN within the retrieval phase. For large $\eta$ (right-hand side), instead, memory of the initial state is quickly lost; the dynamics does not appreciably approach either pattern, implying $m_1 \approx m_2 \approx 0$ and corresponding to a case with a single, trivial BA, analogous to the paramagnetic phase of the HNN.

The emergence of multiple BAs for small $\eta$, in which the configuration tends to fluctuate close to one of the recorded patterns, shows that, in spite of the dissipation, the system is still dynamically capable of retrieving part of the stored information. In order to quantify this very capacity, we consider averages over a number $N_{\mathrm{traj}}$ of trajectories, which we denote by $\overline{O(t)}$ for an observable $O$ at time $t$.
%
%The single trajectory dynamics identifies a change in the BA structure of the model, at varying of the parameter $\eta$. Furthermore, multiple BA seem to emerge at small-$\eta$, this pointing out a similarity with the dynamical behaviour of the HNN within its retrieval phase. For this reason, we focus on the dynamics at small $\eta$, taking into account fluctuations due to different trajectories. Given a generic observable $O$ (i.e.~a function of the configuration), we denote by $\overline{O(t)}$ its average over $N_{\mathrm{traj}}$ trajectories \changer{(What about initial conditions?)}. 
%
Figure \ref{fig.stationary}(b) shows the typical evolution of the average overlaps $\overline{\abs{m_\mu(t)}}$ (grey and black curves) at fixed coupling realization, $\eta = 1$ and $N_{\mathrm{traj}} = 200$, where the absolute value is taken to avoid averaging to zero due to the $\vec{\sigma} \leftrightarrow -\vec{\sigma}$ symmetry of the model. These quantities clearly assume finite values at long times. These values, however, are not indicative of the overlaps that can be actually achieved in any given trajectory. This is because, once the dynamics starts exploring different BAs, at most one overlap will be appreciably different from zero at any given time. We clarify this with an example: say that, at a large time $t$, $p N_{\mathrm{traj}}$ trajectories (with $p < 1$) align with the first pattern giving $\abs{m_1(t)} = m$, whereas the others align with the second one ($m_1(t) \approx 0$). Then, $\overline{\abs{m_1(t)}} \approx p m < m$. To obviate this reduction, we also study the averaged maximum $\mathcal{M}(t)=\overline{ \max(|m_{1}(t)|,|m_{2}(t)|)}$ which constitutes a one-component order parameter symmetric under sign change $\vec{\sigma} \leftrightarrow -\vec{\sigma}$ and pattern permutation ($1 \leftarrow 2$); this symmetrization makes $\mathcal{M}(t)$ a more reliable estimate of the overlap that can be achieved within individual trajectories on either pattern.

\emph{Stationary properties---} 
To estimate the extent of the retrieval regime, we now focus on the stationary (long-time) properties and study $\mathcal{M} (t \to \infty)$ as a function of $\eta$. We additionally account for the fact that this quantity depends on the specific realization of the couplings and is thus a random variable. We thereby denote by $\braket{ \mathcal{M} }_{\mathbf{g}}$ its average over the previously-defined doubly-peaked distribution of the $g_{i\mu}$s and plot it against $\eta$ in Fig.~\ref{fig.stationary}(c).

As discussed in Ref.~\cite{PhysRevResearch.2.013198}, when considering the large and the small $\eta$ limits, effectively thermal regimes can occur. At large $\eta$, an infinite temperature scenario emerges. The finite value taken by the disorder-averaged $\braket{\mathcal{M}}_{\mathbf{g}}$ in this regime is due to the finite size of the system.
%the corresponding infinite-temperature average can be estimated to be $\sqrt{4/(\pi N)} \approx 0.16$ for $N = 50$ (see Supplementary Material \cite{SM}), which is indeed close to the actual values the curve settles to for large $\eta$. 
As $\eta$ is decreased, a crossover emerges towards larger and larger values and for $\eta \approx O(1)$ it seems to reach a maximum of approximately $0.5$, meaning $75 \%$ of the configuration spins are aligned with a pattern. 
%The profile in Fig.~\ref{fig.stationary}(c) shows that, within the analysed range, the role of $\eta$ is analogous to that of the temperature in the HNN. 
In this regime, the error bars get larger as well, implying that the system becomes more sensitive to the specific values taken by the spin-boson couplings. Combining these results with the dynamical ones from the previous section, showing the emergence of several approximate BAs, suggests that the two regimes, $\eta$ large and small, reproduce to an extent the physics of the paramagnetic and retrieval phases of a HNN, with $\eta$ playing the role of an effective temperature. Future investigations at variable system size will shed further light on the features of the crossover.

\emph{Influence of pattern noise---} We recall that, in order to connect the physics of the HNN to our model, we considered the couplings $\vec{g}_{\mu}$ as noisy patterns $\vec{\mathcal{G}}_\mu$. One can thus distinguish between two sources of fluctuations: (i) the randomness of the patterns, whose components can be either $\pm 1$ with equal probability , and (ii) the noise due to the finite width $s$ of the distribution peaks around $\pm 1$. 

It is worth noticing that, while (i) is present in the HNN as well, (ii) represents an additional source of noise specific to our model. For this reason, we wish to better understand its impact on the retrieval ability of the dynamics, reporting the result in  Fig.~\ref{fig.stationary}(d). After having excluded the rightmost points $\eta > 6$, we compare the data used for the curve in panel (c), corresponding to $s = 0.25$ (red) to two additional data sets analogously obtained for $s = 0.125$ (magenta) and $0.063$ (blue). Interestingly, reducing $s$ does not seem to affect the standard deviation (i.e. the error bars); instead, one can spot a decrease in the average values $\braket{\mathcal{M}}_{\mathbf{g}}$. This suggests that small amounts of noise over the patterns can very slightly improve the retrieval of stored information. Although the cause is not entirely clear, it is worth mentioning that a similar effect can be observed in the HNN too (see Ref. \cite{SM} for further details), where, at finite size, the introduction of disorder over the patterns induces a broadening of the order parameter profile. This broadening results in an effective increase of the overlaps close to the critical point; as one moves deeper in the retrieval phase, however, the effect is reversed and the average overlap decreases at larger noise levels.

\emph{Conclusions---} We have investigated the dynamics and the stationary phases of a dissipative multi-mode spin-boson model, highlighting analogies between this many-body system and the HNN, such as a pattern retrieval dynamics and a crossover between a retrieval and a paramagnetic phase. Beyond mimicking the HNN, physical realizations of this spin-boson Hamiltonian, e.g. atom-cavity setups, have the potential to systematically probe the quantum regime, i.e. go beyond the perturbative limit considered here. Increasing the degree of "quantumness", i.e. adding the possibility to host superpositions and entanglement offers intriguing potential for exploring quantum effects in the context of machine learning.

\emph{Acknowledgments---} I.~L.~acknowledges support from EPSRC [Grant No. EP/R04421X/1], from the "Wissenschaftler Rückkehrprogramm GSO/CZS" of the Carl-Zeiss-Stiftung and the German Scholars Organization e.V., and through the Excellence Cluster "Machine Learning: New Perspectives for Science", and through the Deutsche Forschungsgemeinschaft (DFG, German Research Foundation) under Germany's Excellence Strategy - EXC number 2064/1 - Project number 390727645. F.~C.~acknowledges support through a Teach@Tübingen Fellowship. P.~R.~acknowledges funding by the European Union through the H2020 - MCIF No. 766442. Research carried out by M.~M.~for this work was funded by the University of Nottingham through a Nottingham Research Fellowship.

\bibliographystyle{apsrev4-1}

\bibliography{DM_bib}

\widetext
\newpage

\begin{center}
        \textbf{\large Supplemental Material for "Signature of associative memory behavior in multi-mode spin-boson models"}
\end{center}

%%%%%%%%%% Merge with supplemental materials %%%%%%%%%%
%%%%%%%%%% Prefix a "S" to all equations, figures, tables and reset the counter %%%%%%%%%%
\setcounter{equation}{0}
\setcounter{figure}{0}
\setcounter{table}{0}
\makeatletter
\renewcommand{\theequation}{S\arabic{equation}}
\renewcommand{\thefigure}{S\arabic{figure}}

\renewcommand{\bibnumfmt}[1]{[S#1]}

%%%%%%%%%% Prefix a "S" to all equations, figures, tables and reset the counter %%%%%%%%%%

\section{Derivation of the rates}

In this section we provide details on how Eq.(2) of the main text is derived. The state of the system, $\rho(t)$, evolves according to the Lindblad equation $\rho(t)=\lind[\rho(t)]=-i[\hat{H},\rho(t)]+\sum_{n=l,g}\hat{L}_{n}\rho(t)\hat{L}_{n}^{\dagger}-\frac{1}{2}\lbrace \hat{L}_{n}^{\dagger}\hat{L}_{n}, \rho(t) \rbrace$. In order to obtain an effective description of the spin dynamics at strong coupling, we consider the evolution of the state as $\dot{\rho}=\lind_{0}(\rho)+\lind_{1}(\rho)$, with $\lind_{0}=\lind-\lind_{1}$ and $\lind_{1}(\cdot)=-i\Omega[\sum_{i}\hat{\sigma}^{x}_{i}, \cdot ]$. 

We firstly focus on the dynamics generated by $\lind_{0}$. It is worth noticing that each spin operator $\hat{\sigma}_{i}^{z}$ is a conserved quantity under the action of the superoperator $\lind_{0}$. As a result, the eigenstates of $\hat{\sigma}_{i}^{z}$ operators, $\ket{\vs}= \ket{(\sigma_{1}, ..., \sigma_{N})^{T}}$, with $\hat{\sigma}_{i}^{z}\ket{\vs}=\sigma_{i}\ket{\vs}$ and th $\sigma_{i}=\pm 1$, identify $2^{N}$ spin sectors. These are spanned by the combination $\ket{\vs}\bra{\vs}$ and represent $2^{N}$ degenerate stationary subspaces. Thus, we can assume the stationary state of $\lind_{0}$ of the form $\rho_{\mathrm{stat}}=\sum_{\vs}p_{\vs}\rho_{\vs}\ket{\vs}\bra{\vs}$, where $p_{\vs}$ are a set of classical probabilities and $\rho_{\sigma}$ is the corresponding bosonic state, that we assume to be a Gaussian state.  

Secondly, we consider the superoperator $\lind_{1}$, which mixes different spin sectors, as a perturbation with respect to $\lind_{0}$, i.e. we consider $\Omega \ll g$. Indeed, projecting the dynamics onto the stationary manifold of $\lind_{0}$, we exploit the Nakajima-Zwanzig formalism, up to the order $O(\Omega^{2})$, to write the evolution of the spin as 
\begin{equation}\label{s.NZ}
\begin{split}
& P\dot{\rho}^{\mathrm{spin}}(\tau) =  \tr_{B} \int_{0}^{+\infty} d\tau' P \lind_{1}e^{\lind_{0}\tau'}\lind_{1}\rho_{\mathrm{stat}}(\tau)=\\
=& \Omega^{2}\sum_{\lbrace \vs \rbrace}p_{\vs}(\tau)\sum_{i}\int_{0}^{+\infty} d\tau' \sum_{j=\pm} \tr_{B}\left[ e^{\V_{\vs,i}^{j}\tau'}(\rho_{\vs}) \right] \times \\
& \times \left( \hat{\sigma}_{i}^{x}\ket{\vs}\bra{\vs}\hat{\sigma}^{x}_{i} - \ket{\vs}\bra{\vs} \right),
\end{split}
\end{equation}
where $P\dot{\rho}^{\mathrm{spin}}(\tau)=\tr_{B}[\dot{\rho}_{\mathrm{stat}}(\tau)]=\sum_{\vs}\dot{p}_{\vs}(\tau)\ket{\vs}\bra{\vs}$, and $\tr_{B}$ is the partial trace over the bosonic modes. Furthermore, we have defined the spin-configuration dependent superoperators 
\begin{equation}
\V_{\vs,i}^{\pm}(\cdot)= \sum_{\mu}\lbrace - i \omega_{\mu} [\hat{a}_{\mu}^{\dagger}\hat{a}_{\mu},\cdot ] + \dgm \left(\cdot \right)+ \dkm \left(\cdot \right)- iM_{l \mu}\left[(\hat{a}^{\dagger}_{\mu}+\hat{a}_{\mu}), \cdot \right]\pm ig_{l \mu}\sigma_{l} \left\lbrace (\hat{a}^{\dagger}_{\mu}+\hat{a}_{\mu}),\cdot \right\rbrace \rbrace,
\end{equation}
with $M_{l\mu}=\sum_{n\neq l}g_{n \mu}\sigma_{n}$, and $\dgm$, $\dkm$ the dissipative terms representing loss and gain, respectively. By projecting Eq.(\ref{s.NZ}) on a state $\ket{\vs'}$, the dynamics reduces to the evolution of the classical probabilities  ruled by a master equation as
\begin{equation}\label{e.spindynamics}
\dot{p}_{\vs}=\sum_{\vs'} \left( W_{\vs' \rightarrow \vs}p_{\vs'}- W_{\vs \rightarrow \vs'}p_{\vs} \right)
\end{equation}
where $W_{\vs \rightarrow \vs'}=\Omega^2 \int_{0}^{+\infty} d\tau \sum_{j=\pm} \tr_{B}\left[ e^{\V_{\vs,i}^{j}\tau}(\rho_{\vs}) \right]$, is the transition rate for the switching  $\vs \rightarrow \vs'$ and it allows only spin flip processes.   

For going ahead in evaluating the rate expression, we exploit the superoperator properties and we write $e^{\V_{\vs,i}^{\pm}}(\rho_{\vs})=(e^{\V_{\vs,i}^{\pm, *}}\mathbb{I})\rho_{\vs}$, with $\V_{\vs,i}^{\pm, *}$ the adjoint superoperator of $\V_{\vs,i}^{\pm}$ and $\mathbb{I}$ the identity operator. Moreover, the identity operator can be expressed in terms of a generalised displacement operator of a multi-mode bosonic coherent state as $\mathbb{I}=\hat{\mathcal{D}}(0)$, where $\hat{\mathcal{D}}(\tau)=e^{\sum_{\mu} \alpha_{\mu}(\tau) \hat{a}_{\mu}^{\dagger}-\beta_{\mu}(\tau)\hat{a}_{\mu}}e^{-\sum_{\mu}\gamma_{\mu}(\tau)}$ with $\alpha_{\mu}(0)=\beta_{\mu}(0)=\gamma_{\mu}(0)=0$, $\forall \mu$. The operator $\hat{\mathcal{D}}$ can be written as $\hat{\mathcal{D}}(\tau)=\prod_{\mu} \hat{D}_{\mu}(\tau)e^{-\gamma_{\mu}(\tau)}$, where $\hat{D}_{\mu}(\tau)=e^{\alpha_{\mu}(\tau)\hat{a}^{\dagger}_{\mu}-\beta_{\mu}(\tau)\hat{a}_{\mu}}$, due to the commutation of bosonic operators referring to different modes.  We then verify that the displacement operator $\hat{\mathcal{D}}(0)$ is mapped into the generalised one $\hat{\mathcal{D}}(\tau)$ by applying  the adjoint superoperator $\V_{\vs,i}^{\pm}$. To this end, we consider the differential equation $\frac{d}{d\tau}\left[ \hat{\mathcal{D}}_{\vs,i}^{\pm}(\tau)\right]=\V_{\vs,i}^{\pm *}\left[ \hat{\mathcal{D}}_{\vs,i}^{\pm}(\tau) \right]$ and we obtain the solutions for the functions $\alpha^{+}_{\sigma_{i} \mu}(\tau)=\alpha^{-}_{-\sigma_{i} \mu}(\tau)$, $\beta^{+}_{\sigma_{i} \mu}(\tau)=\beta^{-}_{-\sigma_{i} \mu}(\tau)$, $\gamma^{+}_{\sigma_{i} \mu}(\tau)=\gamma^{-}_{-\sigma_{i} \mu}(\tau)$.  For initial conditions $\alpha_{\sigma_{i} \mu}^{\pm}(0)=\beta_{\sigma_{i} \mu}^{\pm}(0)=\gamma_{\sigma_{i} \mu}^{\pm}(0)=0$, and considering, for the sake of simplicity, $\omega_{\mu} \equiv \omega$, $\gamma_{\mu} \equiv \gamma$, $\kappa_{\mu} \equiv \kappa$ $\forall \mu$, we get
\begin{eqnarray}
\begin{split}
& \alpha^{+}_{\sigma_{i} \mu}(t)= [\beta_{\sigma_{i} \mu}^{+}(t)]^{*}=\frac{i4g_{i \mu}\sigma_{i}}{\omega(\eta-2i)}\left[1- e^{(i  - \frac{\eta}{2})t} \right],\\
& \gamma_{\sigma_{i} \mu}^{+}(t)=  \frac{2g_{i \mu}^{2}\nu}{\omega^{2}\eta}\left[f_{1}(t) +t \right] + \frac{ig_{i \mu}\sigma_{i}M_{i \mu}}{\omega^{2}(\eta^{2}+4)}[s(t)+t]\, ,\\
& f_{1}(t)=\frac{1-e^{\eta t}}{\eta}-\frac{4\eta[1-e^{-\frac{\eta}{2}t}\cos{(t)}]-8e^{-\frac{\eta}{2}t}\sin(t)}{\eta^{2}+4},\\
& s(t) = \frac{  4\eta\left[ e^{-\frac{\eta}{2}t}\cos(t)-1 \right] + \left(\eta^{2}-4 \right)e^{-\frac{\eta}{2}t}\sin(t)  }{ \eta^{2}+4}\,,
\end{split}
\end{eqnarray}
where we have defined the dimensionless time $t=\tau \omega$, and $\eta=(\gamma-\kappa)/\omega$,  $\theta = \kappa/\gamma \in [0,1)$, and $\nu=4(1+\theta)\eta/[(\eta^2+4)(1-\theta)]$.

The previous steps allows us to write the partial trace over the boson as $\tr_{B}[e^{\V_{\vs,i}^{\pm}}(\rho_{\vs})]=e^{-\sum_{\mu}\gamma_{\sigma_{i} \mu}^{\pm}(t)}\tr_{B}[\hat{\mathcal{D}}^{\pm}_{\vs,i}(t)\rho_{\vs}] $. Recalling that the bosonic state $\rho_{\vs}$ has been considered to be a multi-mode Gaussian state, we recognize the quantity $\tr_{B}[\hat{\mathcal{D}}^{\pm}_{\vs,i}(t) \rho_{\vs}] $ as the \textit{characteristic function} $\chi^{\pm}_{\vs,i}(t)$ of the state $\rho_{\vs}$. The expression of the characteristic function for a generic Gaussian state $\rho_{G}$ of a $p$-mode bosonic field reads  
\begin{eqnarray}\label{s.chi}
\begin{split}
& \chi_{\rho_{\mathrm{G}}}= e^{\vec{\alpha}^{T} \mathbf{\Omega} \vec{a}_{G}-\frac{1}{4} \vec{\alpha}^{T} \mathbf{\Sigma} \vec{\alpha}} = e^{\sum_{\mu=1}^{p}[\alpha_{\mu}\braket{\hat{a}^{\dagger}}_{G}-\alpha_{\mu}^{*}\braket{\hat{a}}_{G}-\frac{1}{4}\vec{\alpha}_{\mu}^{T}\Sigma_{\mu}\vec{\alpha}_{\mu}]} \\
& \mathbf{\Omega}=\bigoplus_{\mu=1}^{p}\Omega_{\mu}, \, \Omega_{\mu}= \begin{pmatrix}
0 && 1 \\
-1 && 0 
\end{pmatrix}\\
& \mathbf{\Sigma} = \bigoplus_{\mu=1}^{p}\Sigma_{\mu}, \, 
\end{split}
\end{eqnarray}
where $\vec{\alpha}^{T}= (\alpha_{1},\alpha_{1}^{*},...,\alpha_{p},\alpha_{p}^{*})$, $\vec{\alpha}_{\mu}^{T}=(\alpha_{\mu},\alpha_{\mu}^{*})$, $\vec{a}_{G}^{T}=(\braket{\hat{a}_{1}}_{G},\braket{\hat{a}_{1}^{\dagger}}_{G},...,\braket{\hat{a}_{p}}_{G},\braket{\hat{a}_{p}^{\dagger}}_{G}) $, and $\braket{\cdot}_{G}$ are the expectation values of bosonic operators on the Gaussian sate $\rho_{G}$, and $\Sigma_{\mu}$ is the single mode covariance matrix and reads
\begin{equation}\label{s.covariance}
2\begin{pmatrix}
-(\braket{\hat{a}^{2}}_{G}-\braket{\hat{a}}^{2}_{G}) && \frac{1}{2}(\braket{\hat{a}^{\dagger}\hat{a}}_{G}+\braket{\hat{a}\hat{a}^{\dagger}}_{G})-\braket{\hat{a}^{\dagger}}_{G}\braket{\hat{a}}_{G}\\
\frac{1}{2}(\braket{\hat{a}^{\dagger}\hat{a}}_{c}+\braket{\hat{a}\hat{a}^{\dagger}}_{G})-\braket{\hat{a}^{\dagger}}_{G}\braket{\hat{a}}_{G} && -(\braket{\hat{a}^{\dagger \; 2}}_{G}-\braket{\hat{a}^{\dagger}}^{2}_{G})
\end{pmatrix},
\end{equation}
where we omit the bosonic mode label $\mu$ for the sake of a lighter notation.
We apply the definition \eqref{s.chi}, with  the expectation values of the bosonic operators obtained by considering the lindblad operator $\lind_{0}$. The characteristic function reads
\begin{equation}
\chi^{\pm}_{\vs,i}(t)= \exp{\left\lbrace \sum_{\mu} \left[-\frac{1+\theta}{2(1-\theta)}|\alpha_{\sigma_{i} \mu}^{\pm}(t)|^{ 2}+ \frac{2ig_{i \mu}M_{\mu}}{\omega}\left(\frac{\alpha_{\sigma_{i} \mu}^{\pm}(t)}{\eta-2i}+\frac{\alpha_{\sigma_{i} \mu}^{\pm *}(t)}{\eta+2i} \right)\right] \right\rbrace},
\end{equation}
where $M_{\mu}=\sum_{l}g_{l \mu}\sigma_{l}$. Thus, the final expression of the rate is
\begin{equation}
\begin{split}
W_{\vs \rightarrow \vs'}= & \frac{\Omega^2}{\omega} \int_{0}^{+\infty} dt \sum_{j=\pm}e^{-\sum_{\mu}\gamma^{j}_{\sigma_{i} \mu}(t)} \chi_{\vs,i}^{j}(t)= \frac{2\Omega^{2}}{\omega}\int_{0}^{\infty}\! dt e^{-\frac{2g_{i}^{2}\nu}{\omega^{2}}\left[ f(t)+t \right] } \begin{medsize}  \cos{\left[ 16 \frac{\Delta E_{i} t - g_{i}^{2}s(t) }{\omega^{2}(\eta^{2}+4)} \right]} \end{medsize}\,, \\
&f(t) = \frac{-2\eta^2+8}{\eta \left(\eta^2+4\right)} \left[ 1-e^{-\frac{\eta}{2}t}\cos( t) \right]-\frac{8 e^{-\frac{\eta}{2}t} }{\eta^{2}+4}\sin( t)\,, \\
\end{split}
\end{equation}
where $g_{i}^{2} \equiv \sum_{i=1}^{M}g_{i \mu}^{2}$ and the quantity $\Delta E_i~= \sum_{\mu}g_{i \mu} \sigma_i\sum_{l\neq i}g_{l \mu}\sigma_{l}$ retains the dependence on the spin configuration.

\section{Impact of pattern noise in the Hopfield Neural Network}

In this section we compare the behavior of our spin-boson model to that of a HNN with noisy patterns. In particular, we generate the pattern components analogously to the spin-boson coupling, i.e. from a bimodal distribution consisting of the superposition of two Gaussians $\mathcal{P}(\xi)=\frac{1}{2}[\mathcal{N}_{+1,s}(\xi)+\mathcal{N}_{-1,s}(\xi)]$, with $N_{x_{0},s}(x)=(2\pi s)^{1/2}\exp{-(x-x_{0})^{2}/(2s^{2})}$. 
%
%
%given in terms of two gaussians picked at $\pm 1$, as represented in the inset of Fig.(3d) of the main text. In other words, we pick the pattern components $\xi_{i \mu}$ from a bimodal distribution $\mathcal{P}(\xi)=\frac{1}{2}[\mathcal{N}_{+1,s}(\xi)+\mathcal{N}_{-1,s}(\xi)]$ where $N_{x_{0},s}(x)=(2\pi s)^{1/2}\exp{-(x-x_{0})^{2}/(2s^{2})}$. It is worth noticing that in this case fluctuations are due also to the finite width $s$ of the gaussian distributions.

The results, reported in Fig.[S1], have been obtained by means of a Monte Carlo sampling of the HNN's configurations at different temperatures. In panels (a) and (b) of Fig.[S1] we show the HNN disorder averaged order parameter, $\braket{\mathcal{M}}_{\mathbf{\xi}}= \braket{\max_{\mu=1,...,p}\overline{|\zeta_{\mu}|}}_{\mathbf{\xi}}$, as a function of the temperature $T$ for $\mu=1,...,p$ with $p=2$ (two memory case) and a $p=3$ (three memory case), respectively. Different lines correspond to different Gaussian widths, as reported in the legend. As can be seen from Fig.[S1a], near the critical point $T=1$, the order parameter increases with $s$. A similar behavior seems to be present as the number of memories increases, as displayed in Fig.[S1b]. 

\begin{figure}
\includegraphics[scale=0.5]{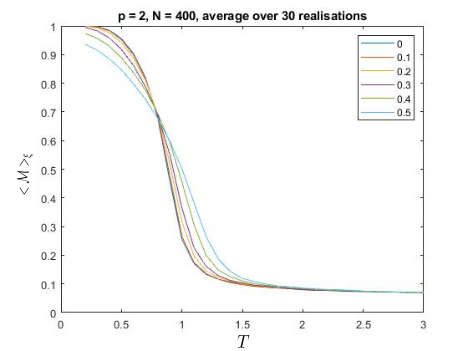}(a)
\includegraphics[scale=0.5]{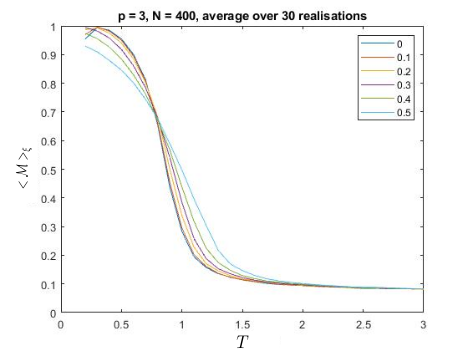}(b)\label{fs.HNN}
\caption{\textbf{Impact of pattern noise on the phase transition in the HNN.} Order parameter as a function of the temperaure. The disorder distribution of the patterns is a bimodal distribution with Gaussian shape, peaked at $\pm 1$. Each line corresponds to a value of the Gaussian width $s$. The two panels report the results in (a) the two memory case, $p=2$, and (b) the three memory case, $p=3$.}
\end{figure}

\end{document}